\def\aa{{A\&A}}
\def\aas{{A\&AS}}
\def\aj{{AJ}}
\def\annrev{{ARA\&A}}
\def\apj{{ApJ}}
\def\apjs{{ApJS}}
\def\baas{{BAAS}}
\def\mnras{{MNRAS}}
\def\nat{{Nature}}
\def\pasp{{PASP}}

\documentclass[cup5b]{caps}
\usepackage{graphicx}
\usepackage{amssymb}
\usepackage{ociwsymp4e}  

\HeadText{P. Molaro et al.  } 

\begin{document}

\pagenumbering{arabic}

\author[]{P. Molaro, M. Centuri\'on, V. D'Odorico, 
 C. P\'eroux\\Osservatorio Astronomico di  Trieste-INAF}

%
%

\chapter{ Nitrogen Abundances in High-z DLAs}

\begin{abstract}
Determination of  chemical abundances for elements
produced mainly by Type I SNae and intermediate mass stars
in high redshift DLAs probes the early chemical build-up
on time-scales    comparable with their  
production.     
Nitrogen shows a peculiar behaviour never detected
before in any other class of objects. 
For [N/H] <~$-3$ there is a plateau with [N/Si]~=~$-1.45$
($\pm $0.05). We interpret this  as   empirical evidence for primary 
N production by massive stars in   young systems where
AGB stars have not yet had time to make their contribution.
The plateau  provides the  observational integrated 
yields for N production by massive stars which are
   theoretically rather uncertain.  
High N/Si and solar [$\alpha$/iron-peak] ratios are
observed at high redshift and place  at an earlier epoch
the onset of star formation. On the other hand, low N/Si,
i.e.  young objects, are observed also at relatively low
redshifts. These evidences suggest that  DLAs started to
be formed at a very early epoch but their formation has
been extended up to later times.
  
\end{abstract}

\section{Introduction}
Damped Lyman-$\alpha$ (DLA) absorption systems  are neutral clouds 
 with large HI column densities  (N(HI)~$\ge 2 \cdot 10^{20} $ cm$^{-2}$).
They are likely protogalactic clumps embedded in dark
 matter halos which are the progenitors of the present
 variety of galaxian populations. 
Highly accurate abundances ($\approx$ 10\%) are derived 
for a variety of chemical elements  in DLAs showing
metallicities  in the range $-2.5<$~[Fe/H]~$<-1.0$  and    
a mild evolution with redshift  (see Prochaska, these proceedings).

Abundances are  measured  up to $z$~=~4.4 or  within 1 Gyr
from the Big Bang (BB) and    less from the reionization
epoch. 
At these redshifts the maximum allowed time elapsed from
the unknown epoch of formation  of the systems and the
observed redshift starts to be comparable with the
nucleosynthesis time-scale for certain elements such as
those produced mainly by Type I SNae or by intermediate
mass stars (IMS). 
In these proceedings, we focus on the abundances of nitrogen and
sulphur  in a sample of high redshift DLAS with the
purpose of re-examining their  [N/$\alpha$] and
[$\alpha$/Fe] ratios. 
More details can be found in Centuri\'on et al (2003) and
D'Odorico and Molaro (2003).

\section{Nitrogen}
 
\begin{figure}
                  \centering
                  \includegraphics[width=8.0cm,height=11cm,angle=-90]{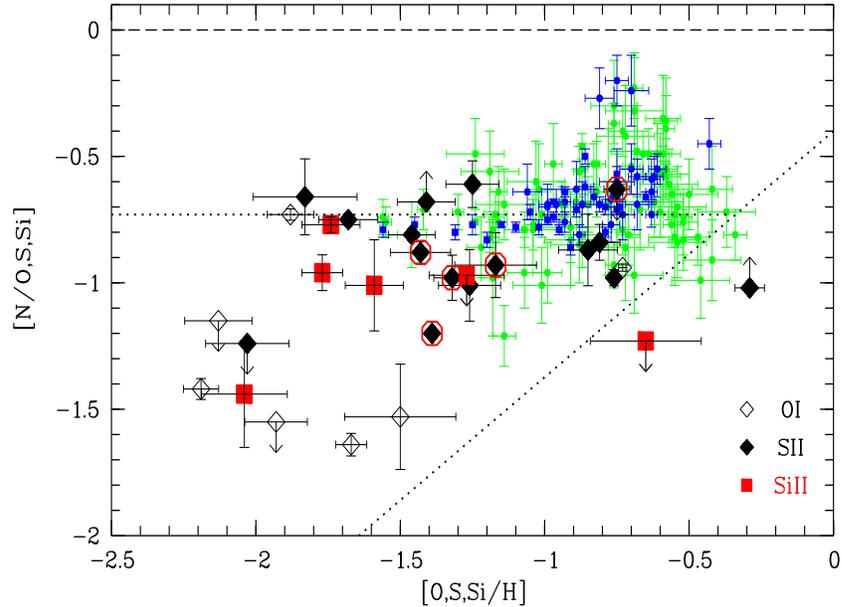}
                                   \caption{[N/$\alpha$] ratio versus
  $\alpha$-element abundance  
  for DLAs and  HII  galaxies (small symbols) from  Centuri\'on et
  al. (2003). } 
                   \label{nalpha}
                \end{figure}   

Nitrogen is thought to be  produced in intermediate mass
stars with masses in the range between 4-8 M$_{\odot}$ which
undergo hot bottom burning in the AGB phase. 
Thus N is restored into the ISM with a  delay which
depends on the lifetime of the  stellar progenitors which
is significantly larger than that of the  Type II products.                 
According to Henry et al. (2000) the bulk of N production
occurs   after $\sim 250-300$ Myr.

Figure~\ref{nalpha} shows the behaviour of  [N/$\alpha$] ratio versus 
$\alpha$-element abundances for a compilation of  DLAs
from Centuri\'on et al. (2003) compared with those of 
HII  galaxies.  The elements   representative of the
$\alpha$-element abundances are  in order of priority:
oxygen,  sulphur and   silicon. 
The figure shows  the  presence of a large scatter in the
distribution of the N/$\alpha$ points with several DLAs
below the extragalactic HII regions.  
Prochaska et al. (2002)  recently claimed the presence of
a bimodal distribution in the  [N/Si]-[Si/H] plane, based
on   high quality Keck observations of two objects
found to have  similar [N/Si], much lower than  the
majority of the other DLA systems.  
There are growing evidences of the existence of such a
group with low N/Si values.    
Very low values are observed  towards APM BR J0307-4945  
(Dessauges-Zavadsky et al. 2001) and  towards Q2059-360
(Centuri\'on et al. 2003). Pettini et al. (2002) provided
three stringent limits for N/Si. 
Thus we have  a total of  4 determinations and 3 upper limits
which look disconnected from the bulk of the DLAs.

\section{The case of  Q2206-199}

\begin{figure}
                  \centering
                  \includegraphics[width=8.0cm,height=11cm,angle=-90]{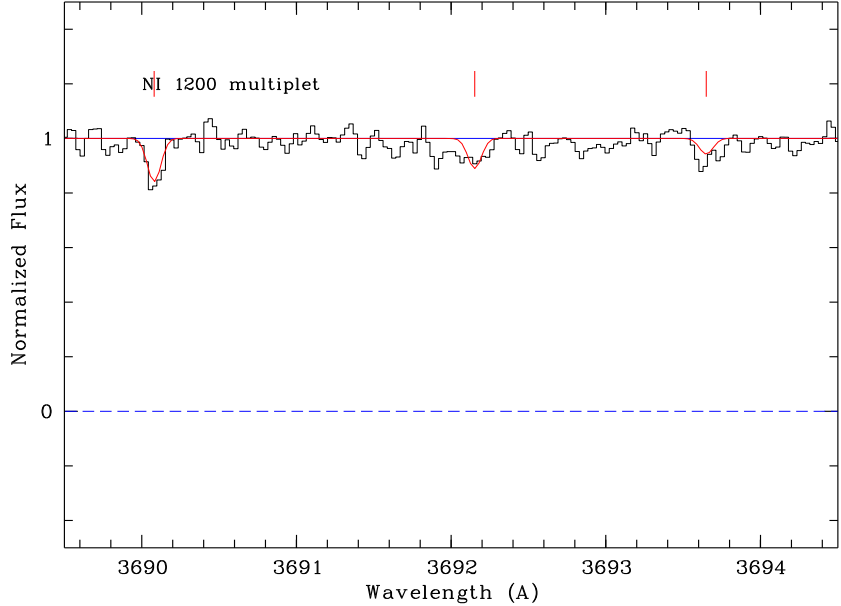}
                  \caption{ The NI 1200 \AA~ multiplet of the DLA at 
                  z$_{abs}$=2.0762
                  towards Q2206-199.}
                   \label{n_2206}
                \end{figure}

The DLA at $z_{\rm abs}$~=~2.07622 towards Q2206-199 is one
of the 3 upper limits studied by Pettini et al. (2002).   
Inspection of  their fit  for the NI 1200 \AA~ multiplet 
shows some inconsistency between the two stronger transitions.
From their Fig.~5, one notes that the
abundance which reproduces the NI  1199.5 \AA~ line is too
strong to fit the NI 1200.2 \AA~ transition line. 
However, a shallow absorption is present at the base of
the stronger feature. This relatively broad absorption
can be produced by a HI interloper with a column density
of log N(HI)~=~12.95~($\pm$ 0.02) and a broadening of
$\it b$~=~40 km s$^{-1}$. 
A new N determination which accounts for  the  HI
contamination corresponds to log~N(NI)~=~12.58~($\pm
0.05$) and $\it b$~=~3 km s$^{-1}$ and is   shown in
Fig.~\ref{n_2206}.  When we combine it  with the Si measured in the
system we obtain  [N/Si]~=~$-1.44~(\pm 0.05)$. 
Thus also this system is found to fall precisely onto  the low N/Si 
 plateau which is now represented  by five systems.

 \section{A low-[N/Si] plateau: existence and possible explanation}

\begin{figure}
               \centering
               \includegraphics[width=7.5cm,height=12cm,angle=-90]{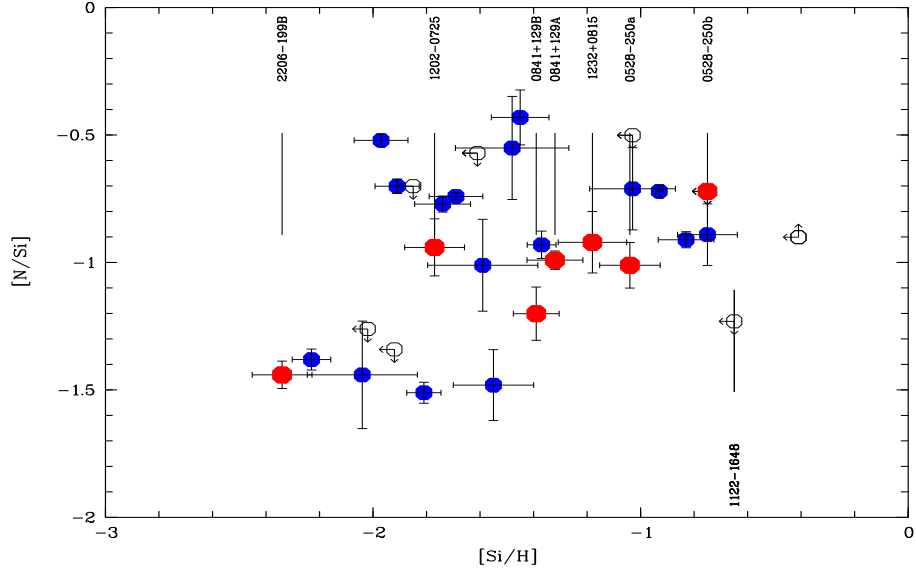}
                  \caption{[N/Si] ratio versus [Si/H] abundance 
  for DLAs. Red  points (lighter  ones   in B\&W)  with the label 
  are the new measurements. Empty circles are limits.  }
                   \label{n_si}
\end{figure}

\begin{figure}
             \centering
             \includegraphics[width=7.5cm,height=12cm,angle=-90]{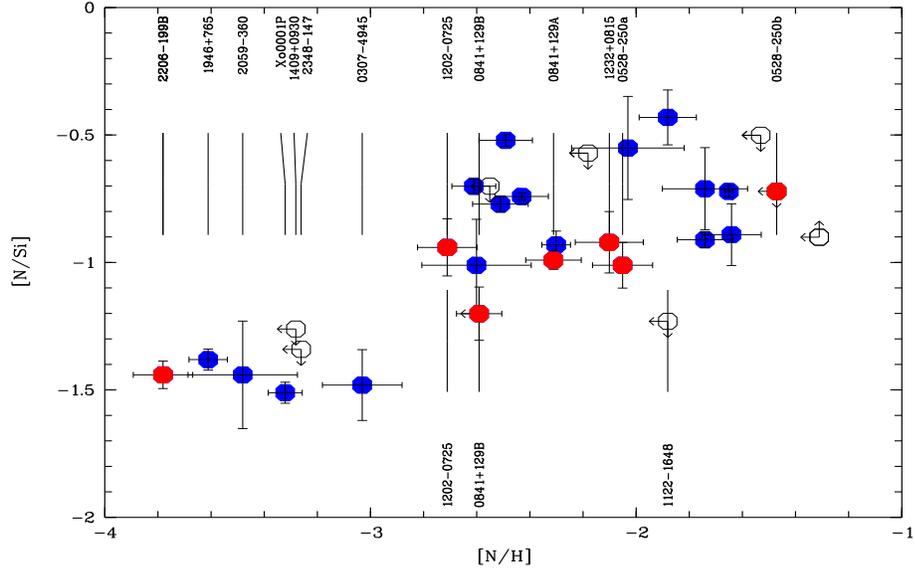} 
             \caption{[N/$\alpha$] ratio versus nitrogen abundance 
  for DLAs. Red  points (lighter  ones   in B\&W)  are the new measurements.
   }          \label{n_acca}
\end{figure}

   In   Fig.~\ref{n_si}  all the data points are taken
with respect to Si, which  although refractory   is only
mildly depleted and has the advantage of being measured
in almost all the DLAs.  
At face value, it is not clear from Fig.~\ref{n_si} whether the
bimodal distribution is real or if it is just the result
of the small number  of observations. 
Moreover, the presence of some DLAs with intermediate
N/Si values such as the case towards Q0841+129B which lies
between the two groups suggests some connection between
the  low and high-N/Si groups. 
However, a different  way of plotting the same data    
reveals that there are two really distinct groups with
different evolutionary behaviour. Following Molaro (2003) we plot in
   Fig.~\ref{n_acca}  the N/Si ratio   versus nitrogen
enrichment.       
The plot clearly shows   two different  regimes below and
above approximately [N/H]~$\approx$~$-3$.  
For  [N/H]~<~$-3$ there are all the  low N/Si values
with [N/Si]~=~$-1.446~(\pm 0.025)$, while for
[N/H]~>~$-3$ there are  17 values providing a weighted
mean of [N/Si]~=~$-0.75~(\pm 0.17$). 
A Montecarlo analysis of the errors of   the high N/Si
plateau  gives an  expected dispersion  of 0.10
($\pm$0.02)  thus providing  some evidence for an  
intrinsic dispersion. For the low N/Si plateau the same
analysis gives  an expected dispersion which is
comparable to and even smaller than the observed one
providing no  evidence  for   dispersion.
  
What is the reason for the different  aspect  of the
two plots of Figures~\ref{n_si} and \ref{n_acca} ?  
The $\alpha$-enrichment depends critically on the SFR and
it is plausible that the DLAs have different SF histories. 
This means that they can reach the same amount of
$\alpha$ enrichment at  different times. 
On the other hand, N   production  depends more critically on the 
lifetime of its  progenitors rather than on the SFR. 
Thus plotting the N/Si versus the nitrogen enrichment 
is   closer to a true temporal evolution and the
degeneracy that we see in the 
N/Si-$\alpha$/H plane is resolved in the N/Si-N/H plane.

Low N/Si values have been generally ascribed to young
systems and high N/Si to  DLAs in a more 
evolved status.  However, the presence of a plateau is
not easy to reconcile with this interpretation,  
since we expect a distribution rather than a plateau.
       
A top-heavy  or  truncated IMF  has been suggested 
 by Prochaska et al. (2002) to explain the
low N/Si plateau. 
 However,  N in the low N/Si plateau is found to increase
 in lockstep with the $\alpha$-elements  by one order of
 magnitude, with no detectable  dispersion.
We thus  interpret the low N/Si plateau  as an evidence 
for primary N production by massive stars. 
If the  systems   are younger than the characteristic
 time-scale  for the N production by AGB stars then there
 is no need to invoke any change in the  initial IMF. 
The high  N/Si values   result from the  AGB N production
 in relatively older systems. 
With this interpretation we do not expect to find any object below 
the low N/Si plateau, but some    in between the two
plateaux may be present. 
Indeed,   Q0841+129B with an intermediate N/Si value could be 
one of these cases.

  \begin{figure}
               \centering
               \includegraphics[width=8.0cm,height=11cm,angle=-90]{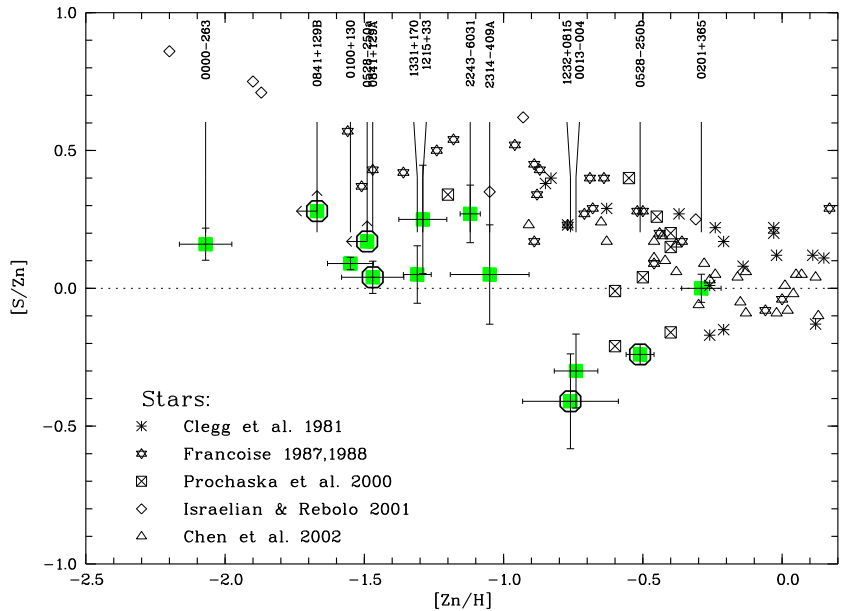}
               \caption{ [S/Zn] ratios. Circles highlight the new measures from
                   Centuri\'on et al. (2003)} 
                                     \label{alpha_zn}
                \end{figure}

Theoretically, N is not expected to be produced in
standard models  of massive stars unless  C in the He-shell mixes with the
H burning shell (Heger \& Woosley 2002).
Rotation  can induce this mixing leading to N  production
(Meynet  and Maeder 2002). The prediction for massive stars (8-120
M$_{\odot}$) with v$\sin$~i~=~400 km s$^{-1}$ and metallicity
Z=1$\cdot$10$^{-5}$ solar is [N/O]=$-$1.5 and about 0.7 dex
below the full mass production (Meynet et al. 2003). 
Also the PopIII yields for  massive stars of Limongi and
Chieffi (2002), computed adopting a high
$^{14}$C($\alpha$,$\gamma$)$^{16}$O rate (Caughlan et al.
1985) can make N. These models coupled with those for the
IMS from Chieffi et al. (2002) computed with a   mass
loss parameter $\eta$~=~6 are able to match the
observations. In fact
the integration with a Salpeter IMF over the full mass
spectrum (4-80 M$_{\odot}$) gives [N/O]$\approx$$-0.85$, while the
integration over only the more massive stars (15-80
M$_{\odot}$) gives [N/O]$\approx$ $-1.5$.
 
We would like to emphasize that for the first time the
observations of DLAs offer  empirical  guidelines to the
theoretical yields  for nitrogen production in massive
stars.

\section{The [$\alpha$/Fe]    ratios}

The presence of an [$\alpha$/Fe] enrichment in  DLAs has
recently been the subject of a debate.   
The problem is that the presence of dust in DLAs
complicates  the interpretation since it  depletes  the
elements differentially. 
When dust has been corrected for or when the
interpretation relies on non-refractory elements only,
the DLAs show a moderate or even absent [$\alpha$/Fe]
enhancement. The new S observations highlighted  with circles in
Fig.~\ref{alpha_zn}  provide new evidence for solar and even subsolar
ratios in the DLAs.

Solar [$\alpha$/Fe] ratios  imply  that most
DLAs are old enough to cause  the Type Ia   to have   
already mixed their products in the medium and have
lowered  the $\alpha$-Fe ratios significantly. 

\begin{figure}
                  \centering
                  \includegraphics[width=8.0cm,height=11cm,angle=-90]{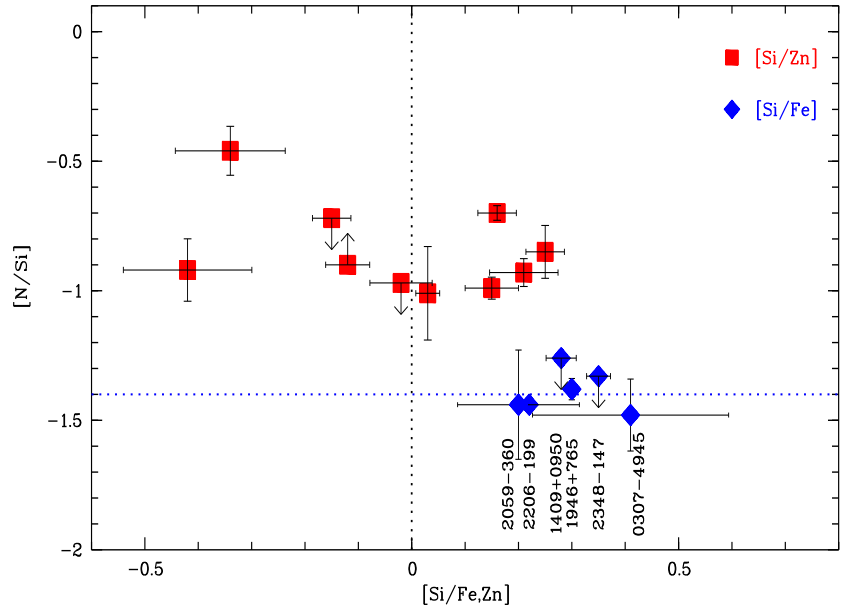}
                  \caption{ [N/Si] ratios versus [Si/Fe] 
                  for diamons  or [Si/Fe] squares.     }
                  \label{n_zn}
                \end{figure}

It is rather interesting to study  the $\alpha$ over iron
ratios in the systems with low N/Si. If these  are
young systems, as we propose,  we should see a marked
[$\alpha$/Fe] enhancement. 
The [Si/Fe] shown in Fig.~\ref{n_zn} are enhanced by about 0.3 dex, almost
in line 
with the other DLAs. Unfortunately, Zn is not measured
thus  preventing an assessment of the presence of dust. 
The case of Q0841+129B  is rather interesting. As we
mentioned before, this system looks like a transition
object and its  age should be very  close to the
characteristic time for N  enrichment by AGBs. 
Thus, it  is rather encouraging to find that
[Si/Zn]~>~0.35 providing in this case a hint of
genuine enrichment as expected, which remains rather unique so far.

 \section{The epoch of DLA formation}

\begin{figure}
                  \centering
                  \includegraphics[width=8.0cm,height=11cm,angle=-90]{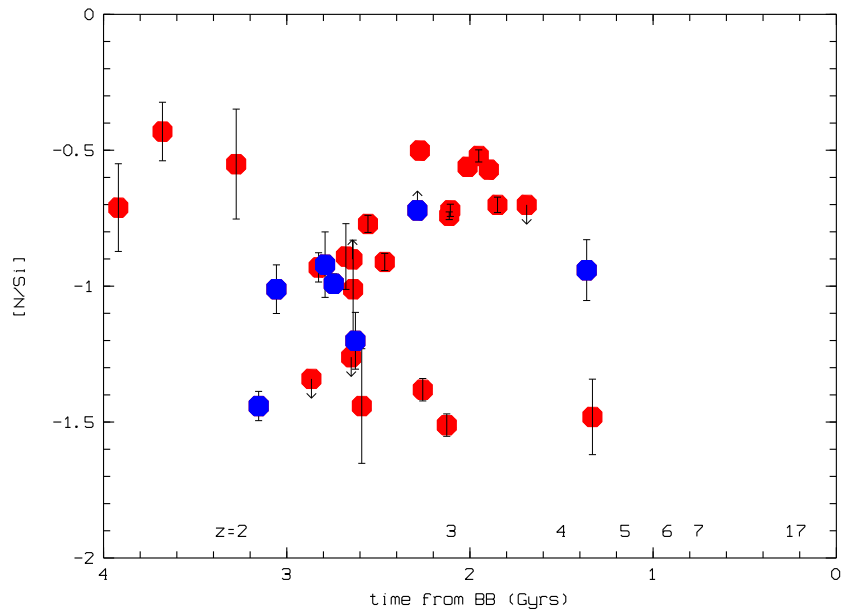}
                  \caption{ [N/Si] ratios versus time.
                   Blue points (darker ones   in B\&W)  are the new
                   measurements } 
                  \label{n_time}
                \end{figure}

Statistically, the fraction of low-N/Si DLAs is  about
25\% of the total, counting also the 2 upper limits. 
This is consistent with a picture of young systems
younger than 250 Myrs and old systems older than about 1
Gyr on average, or more if the time for the full N
release is greater than 250 Myrs. 
In Fig.~\ref{n_time} the N/Si values are plotted versus time since
BB (or redshift) according to present-day cosmology.   
It is possible to see that high and low N/Si are observed
at any time in the range spanned by the observations. 
If we consider 250 Myrs as a representative time for the full  
release  of N   by IMS (it is longer for rotational
models) a high N/Si value such as the one observed
towards Q1202-0725 at $z= 4.4$ put the onset of star
formation at $z$~>~6 or less than 1 Gyr from the BB. 
A similar indication comes from the observation of solar 
[$\alpha$/Fe] ratios in high redshift DLAs, when we
consider the evolutionary time-scales for the Type Ia.
On the other hand, a low N/Si  value such as the one observed 
towards Q2206-199 at $z=2.0$ implies that the 
star formation in this system took place at $z$~<~2.5. 
These results indicate a continuous formation of
DLAs rather than a specific epoch for their formation.

\section{Acknowledgements}

It is a plasure to acknowledge Max Pettini for sharing with us his reduced 
data on QSO Q2206-199.
I would also like to warmly ackowledge Piercarlo Bonifacio for performing
the Montecarlo analysis of the erros and Marco Limongi 
and Alessandro Chieffi for illuminating discussions on the nitrogen 
stellar nucleosynthesis.     
 
%

\def\aa{{A\&A}}
\def\aas{{A\&AS}}
\def\aj{{AJ}}
\def\annrev{{ARA\&A}}
\def\apj{{ApJ}}
\def\apjs{{ApJS}}
\def\baas{{BAAS}}
\def\mnras{{MNRAS}}
\def\nat{{Nature}}
\def\pasp{{PASP}}


\begin{thereferences}{}
 
\bibitem{} Centuri\'on, M., Molaro, P., Vladilo, G.,
Peroux, C., Levshakov, S. A., D'Odorico, V., 2003, A\&A,
403, 55, astro-ph/0302032 

\bibitem{} Caughlan, G.~R., Fowler, W.~A., Harris,
M.~J. Zimmerman, B.~A. 1985, Atomic Data and Nuclear Data
Tables, 32, 197 

\bibitem{} Chieffi, A., Limongi, M., Dominguez, I.,
Straniero, O. 2002, in Chemical Enrichment of Intracluster and
Intergalactic Medium, eds. R. Fusco-Fermiano and F. Matteucci
(ASP Conference Proc.), 253

\bibitem{} Dessauges-Zavadsky, M., D'Odorico, S.,
McMahon, R.~G., Molaro, P., Ledoux, C., P\'eroux, C.,
Storrie-Lombardi, L.~J. 2001, A\&A, 370, 426  

\bibitem{} D'Odorico, V. \& Molaro P.  2003 in preparation

\bibitem{} Heger, A. \& Woosley, S. E. 2002, ApJ, 567, 532

\bibitem{} Henry, R.B.C., Edmunds, M. G. and K\"oppen, J. 2000, Apj, 541, 660.

\bibitem{} Limongi M., and Chieffi A., 2002 PASA, 19, 246

\bibitem{} Meynet,G. \& Maeder, A. 2002, A\&A, 390, 561 

\bibitem{} Meynet, G. et al. astro-ph/0301288

\bibitem{} Molaro, P. 2003, in CNO in the universe 
 C. Charbonnel, D. Schaerer \& G. Meynet eds (ASP Conference Series),
 astro-ph/0301407 

\bibitem{} Pettini, M., Ellison, S.~L., Bergeron, J.,
Petitjean, P. 2002, A\&A, 391, 21 

\bibitem{} Prochaska, J., Henry, R.~B.~C., O'Meara,
J.~M., Tytler, D., Wolfe, A.~M., Kirkman, D., Lubin, D.,
Suzuki, N. 2002, PASP, 114, 933


\end{thereferences}
\end{document}